# Bulk and Two-dimensional Silver and Copper Monohalides: A Unique Class of Materials with Modest Ionicity/Covalency and Ferroelasticity/Multiferroicity

Yaxin Gao[a], Menghao Wu[a]*, Xiao Cheng Zeng[b]*

[a]School of Physics and Wuhan National High Magnetic Field Center, Huazhong University of Science and Technology, Hubei, Wuhan, China 430074. Email: wmh1987@hust.edu.cn

[b]Department of Chemistry and Department of Mechanical & Materials Engineering, University of Nebraska— Lincoln, Lincoln, Nebraska 68588, United States. Email: xzeng1@unl.edu

**Abstract**  Silver and copper monohalides can be viewed as a class of compounds in the "neutral zone" between predominantly covalent and ionic compounds, thereby exhibiting neither strong ionicity nor strong covalency. We show *ab initio* calculation evidence that silver and copper monohalides entail relatively low transition barriers between the *non-polar rock-salt* phase and the *polar zinc-blende* phase, due largely to their unique chemical nature of modest iconicity/covalency. Notably, the low transition barriers endow both monohalides with novel mechanical and electronic properties, i.e., coupled ferroelasticity and ferroelectricity with large polarizations and relatively low switching barriers at ambient conditions. Several halides even possess very similar lattice constants and structures as the prevailing semiconductors such as silicon, thereby enabling epitaxial growth on silicon. Moreover, based on extensive structural search, we find that the most stable two-dimensional (2D) polymorphs of the monolayer halides are close or even greater in energy than their bulk counterparts, a feature not usually seen in the family of rock-salt or zinc-blende semiconductors. The low transition barrier between zinc-blende phase and 2D phase is predicted. Moreover, several 2D monolayer halides also exhibit multiferroicity with coupled ferroelasticity/ferroelectricity, thereby rendering their potential applications as high-density integrated memories for efficient data reading and writing. Their surfaces, covered by halides, also provide oxidation resistance and give low cleave energy from layered structure, suggesting high likelihood of experimental synthesis of these 2D polymorphs.

**Introduction**

Silver halides like AgBr and AgI are well known as photosensitive materials for photographics[1] and photochemistry.[2] They exhibit excellent photocatalytic performance under visible-light irradiation.[3] Their structural, transport, and dynamic properties have also been studied extensively.[4-6] At ambient conditions, bulk silver chloride and silver bromide form a rock-salt (RS) structure, while silver iodide crystallizes in a mixed phase of wurtzite (WZ) and zinc-blende (ZB); it also becomes a super-ionic conductor beyond 148 °C, where the silver ions migrate into the interstitial sites of the bcc sub-lattice of the immobile iodine ions.[7] Due to the small energy difference, phase transition from WZ to ZB may occur upon a relatively modest pressure of 0.1 GPa, and the phase transition to RS structure can take place upon further compression, together with the associated increase in cation–anion coordination from tetrahedral to octahedral.[8-11] Generally, the RS structure tends to exhibit ionic bonding features, whereas the ZB structure tends to exhibit more covalent bonding features.[4] At ambient conditions, the compounds like silver monohalides - copper monohalides (e.g., CuCl, CuBr) - also incline to form the ZB structure.[12] Their ionicity approaches 0.7 according to the Philips scale, slightly lower than the critical value 0.785, marking the idealized boundary between predominantly ''covalent'' and ''ionic'' systems.[13] In this work, we show *ab initio* computation evidence, for the first time, of the possible ferroelectricity[14] and ferroelasticity[15-19] at ambient conditions for the silver and copper monohalides.

It is known that ferroic (ferromagnetic, ferroelectric, or ferroelastic) materials can be applied as non-volatile access memories (RAMs) as these materials possess bi-stable states with degenerate in energy.[20] Indeed, these functional materials provide an alternative approach to overcome some major issues in current Si-based RAMs, e.g., the quantum-tunneling due to the inequivalent bi-states, and power dissipation due to volatile memories. Multiferroic materials with more than one ferroic order parameters may even have the merits of combining different ferroics for efficient data reading and writing.[20, 21] Note that although many materials possess spontaneous polarization/strain, they are not necessarily ferroelectric/ferroelastic, unless a switching pathway entails a relatively low barrier.[22, 23] For example, most WZ and ZB semiconductors like ZnO are polar, due to alternating planes of cations and anions, but are still non-ferroelectric materials due to high switching barrier. For many ZB silver and copper halides considered, both the ferroelectric and ferroelastic switching barriers are lower than 0.15 eV, thereby becoming multiferroic. Moreover, their ferroelectricity and ferroelasticity are coupled, which is required for efficient data reading and writing, since a 90 degree lattice rotation can be driven by either an electric field or a strain, equivalent to a ferroelastic switching and 90-degree ferroelectric switching. Note also that silver and copper halides may be integrated with Si-based circuits via epitaxial growth due to similar lattice constants and structures as Si. The small lattice mismatch may resolve a major issue in integrating traditional ferroelectrics into silicon-based circuits.

Finally, we show that these halides also possess stable two-dimensional (2D) van der Waals polymorphs, and their cohesive energy difference with respect to the bulk phases are within several tenth of meV/f.u.. These 2D monolayers are predicted to be 2D multiferroics as well, with coupled ferroelectricity and ferroelasticity and even lower switching barrier. Such 2D multiferroics with atomic thickness may be more demanded as high-density integrated memories since van der Waals interface does not require lattice matching.

**Computational methods**

Our *ab initio* calculations are performed within the framework of spin-unrestricted density-functional-theory (DFT), implemented in the Vienna *ab initio* Simulation Package (VASP 5.4)[24]. The projector augmented wave (PAW) potentials[25] for the core, and either the local density approximation (LDA) or the generalized gradient approximation (GGA) in the Perdew-Burke-Ernzerhof (PBE)[26] form for the exchange-correlation functional, is applied. The Monkhorst-Pack *k*-meshes are set to 14 × 14 × 10 in the Brillouin zone, and the electron wave function is expanded on a plane-wave basis set with a cutoff energy of 520 eV. All atoms are relaxed in each optimization cycle until atomic forces on each atom are less than 0.01 eV Å$^{-1}$ and the energy variation between subsequent iterations falls below 10$^{-6}$ eV. The Berry phase method is adopted in computing the ferroelectric polarizations[27], and a generalized solid-state elastic band (G-SSNEB) method is used to calculate the pathway of phase transition.[28] An unbiased swarm-intelligence structural method implemented in the CALYPSO code[29, 30] is employed to search for stable 2D silver/copper monohalides.

**Results and discussion**

I.    Phase transition

The RS and ZB structures of silver halide crystals are displayed in Fig. 1(a). Our DFT calculations based on the LDA functional shows that RS structures are the ground state for both AgCl and AgBr, consistent with previous experimental observations. Meanwhile, the ZB structure is energetically more favorable for AgI, as for CuCl and CuBr. Meanwhile, the ZB structure is also predicted to be the ground state for AgCl and AgBr based on the GGA functional, inconsistent with previous experimental reports.[4] Hence, the LDA computation appears to be more reasonable for predicting the ground state structure for this class of materials. In any case, the energy difference between the two phases, $\Delta E=E(ZB)-E(RS)$, is all within 0.09 eV/f.u. for the silver and copper monohalides MX (see Table 1). The absolute values are much smaller, compared with either more ionic compounds such as NaCl or more covalent compounds such as ZnS. Fig. 1(c) plotted $\Delta E$ vs Philips iconicity of the binary semiconductors listed in Table 1, revealing an approximately linear growth model, where the silver and copper monohalides are located within the transition zone marked by the circled region.

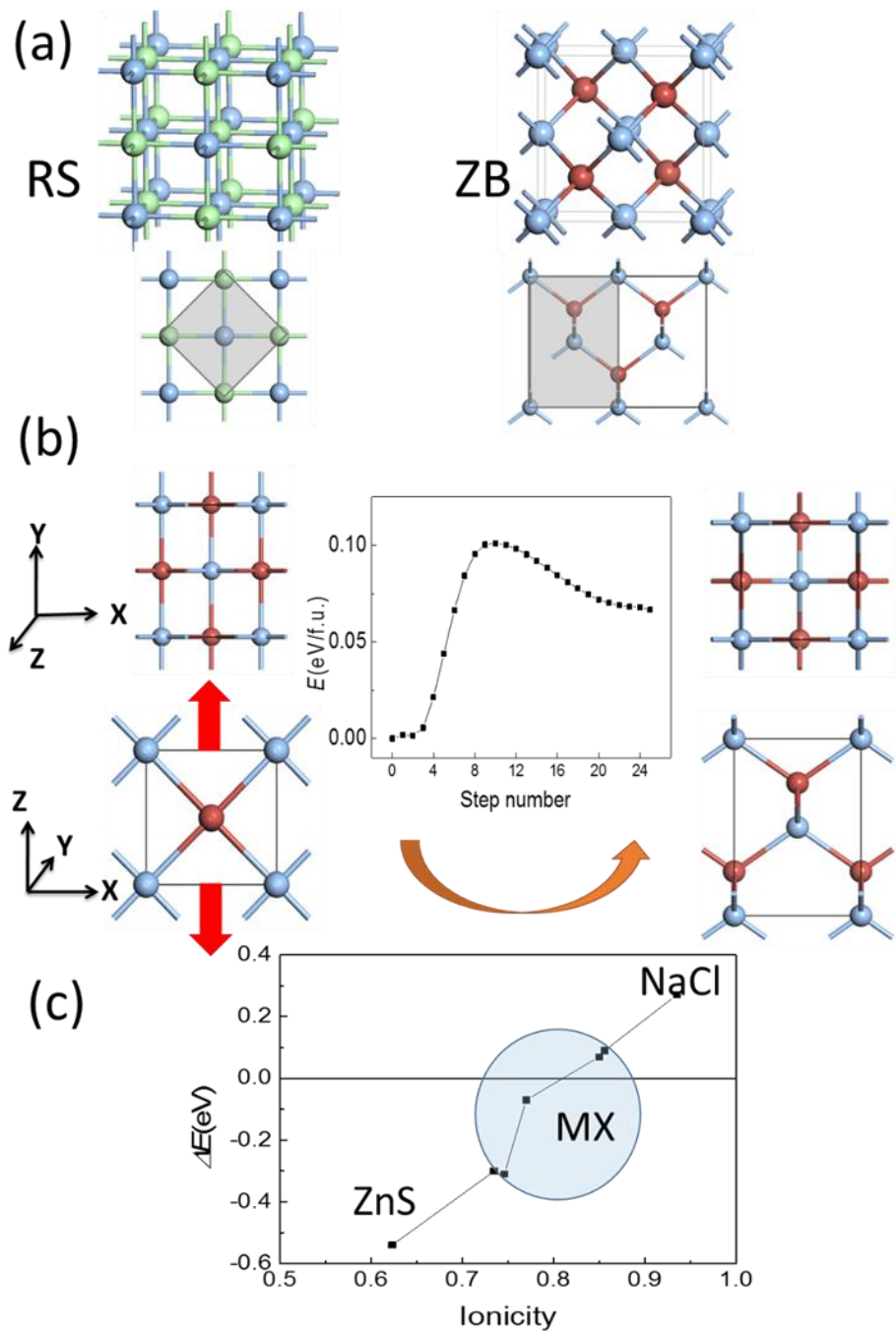

Figure 1. Geometric structure of (a) RS and ZB silver halide crystals, and smaller unit cells marked by the dark area can be adopted. (b) Pathway of phase transition from RS (left) to ZB (right) structure. Blue, green, brown spheres denote Ag, Cl, Br/I atoms, and red arrows denote the strain direction. (c) Energy difference between RS and ZB phase (LDA computation) vs iconicity in Table 1.

The ZB structure can be obtained by simultaneously displacing all halide anions in the RS structure to (¼, ¼, ¼), and the calculated pathway based on SSNEB (see Fig. S1) shows a transition barrier of 0.17 eV/f.u. for AgBr. However, as we rotate the lattice of RS and ZB structures by 45 degree and adopt a smaller unit cell illustrated by the grey area in Fig. 1(a), the ZB structure can be also obtained by buckling the linear –M-X- chains in the RS structure. In view of the distinct lattice parameters and the small energy difference of the two phases, the phase transformation from RS to ZB can be induced by a tensile strain along the z axis by prolonging the lattice constant along -z with the vertical displacement of halide ions and buckling of –M-X- chains, as displayed in Fig. 1(b). For example, the calculated transition barrier is only 0.10 eV/f.u. for AgBr, where the ZB phase, only 0.07 eV/f.u. higher in energy, may become more stable upon a tensile strain. The ZB metastable phase might also be obtained via epitaxial growth on substrates with similar ZB structure and similar lattice constants, e.g., AgCl on Ge, or AgBr on InAs. However, compared with a tensile strain, which can be difficult to control uniformly, an external electric field may be a more feasible approach for inducing the phase transformation, noting that the ZB structure is polar (with alternating planes of cations and anions) while the RS structure is non-polar.

Table 1. Philips ionicity (from Ref. 14) and energy difference between RS and ZB phase calculated by using LDA/GGA (negative values reveal that ZB phase is more favorable in energy).

|  |  | AgCl | AgBr | AgI | CuCl | CuBr | NaCl | ZnS |
|---|---|---|---|---|---|---|---|---|
| **ionicity** |  | 0.856 | 0.850 | 0.770 | 0.746 | 0.735 | 0.935 | 0.623 |
| **Δ$E$(eV/f.u.)** | LDA | 0.09 | 0.07 | -0.07 | -0.31 | -0.30 | 0.27 | -0.54 |
|  | GGA | -0.05 | -0.08 | -0.19 | -0.35 | -0.36 | 0.14 | -0.63 |

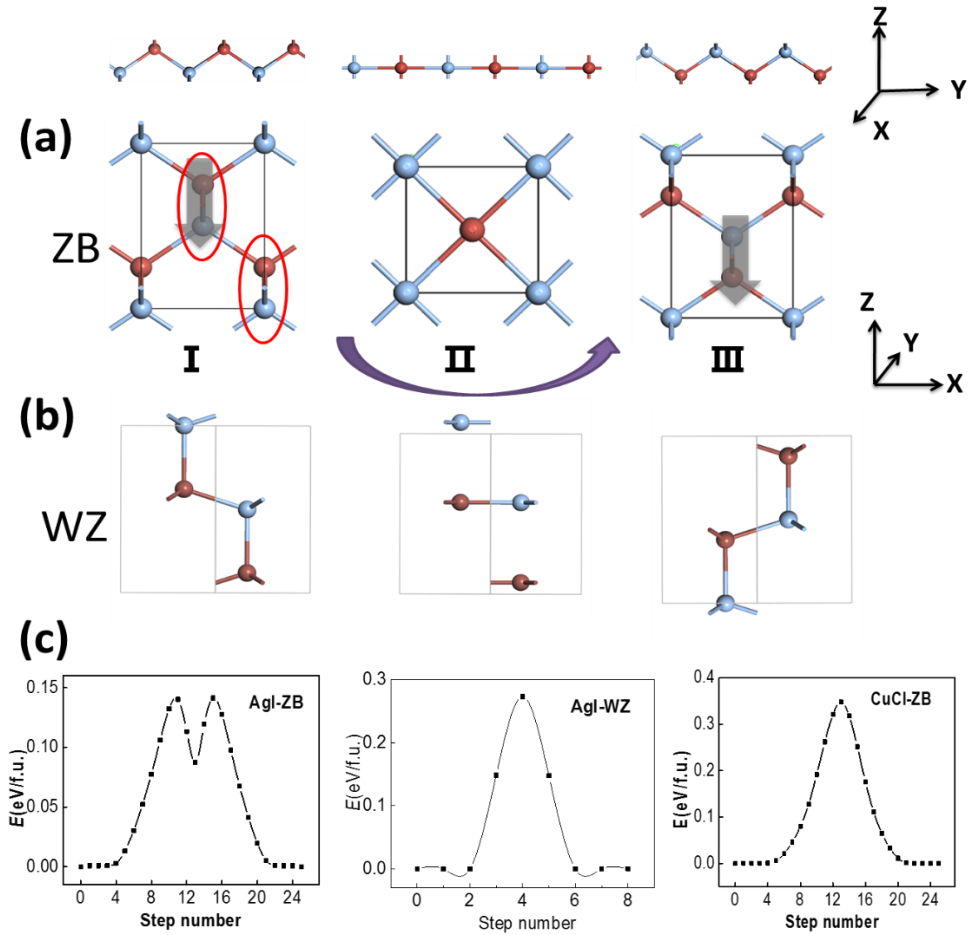

Figure 2. An illustration of the ferroelectric switching pathway of (a) ZB and (b) WZ AgI, where the top panels show that the zigzag chains in red circles (lower panels) at the initial state (I) become flat at the intermediate state (II), and are finally reversed in state III. Thick/grey arrows in (a) denote the direction of polarizations. (c) Calculated energy profile of switching pathway for ZB AgI, WZ AgI and ZB CuCl by using the NEB method.

II. Ferroelasticity/Multiferroicity

The ZB structure is non-centrosymmetric and thus a polarization might be spontaneously formed. The ZB structure of AgI can be viewed as bundles of AgI zigzag chains with their polarization along the z axis, as marked by red circles in Fig. 2(a), and the polarization may be switchable due to a relatively low switching barrier for those chains. We note that the WZ ZnO also possesses a polar structure, but its polarization is not deemed as switchable due to a relatively high switching barrier of 0.25 eV/f.u..[31] It only becomes ferroelectric in the form of ultra-thin multilayers as the associated barrier is lowered to 0.10 eV/f.u. due to the effect of depolarization field.[32] If this polarization can be reversed with the switching of AgI zigzag chains in a possible switching pathway from I to II (see Fig. 2(a)), the symmetrical RS phase can be the intermediate state (II) as the zigzag AgI chains become flat with the halide anions moving along y direction. The switching pathway is computed by using the SSNEB method as shown in Fig. 2(a), suggesting a low switching barrier of 0.10 eV/f.u. The latter is much lower than the switching barrier of PbTiO$_3$ (~0.20eV/f.u.). Meanwhile, a large polarization of 0.40 C/m$^2$ is formed in BZ-AgI, based on our Berry-phase calculation, a value much greater than that in BaTiO$_3$ (~0.26 C/m$^2$). The WZ phase of AgI (0.2meV/f.u.) is also polar with a polarization of 0.20 C/m$^2$, while our SSNEB calculation of the pathway in Fig. 2(c) shows that the ferroelectric switching barrier can be as high as 0.27 eV/f.u., even higher than that for WZ ZnO (0.25 eV), making ferroelectric switching almost impossible at ambient condition.

Similarly, for ZB CuCl, the computed switching pathway shows an even higher barrier of 0. 35eV/f.u., despite the calculated polarization is as high as 0.59 C/m$^2$. However, this barrier is not necessarily the lowest one among all the possible switching pathways, which will be demonstrated later. Compared with AgI, the energy difference between ZB and RS state of CuCl is much larger. The switching pathway of ZB AgI, with a local minima and two transition states, indicates that RS is metastable, whereas for CuCl the RS state is the transition state.

Another possible switching pathway is also taken into consideration. As displayed in Fig. 3(a), the polar zigzag chains in the initial state I may also be switched from –z to the -x or -y axis upon a horizontal electric-field, leading to a 90 degree polarization switching with a lattice rotation at the final state III. Here, the zigzag chains are not be flat like in Fig. 2(a), but can be switched by 90 degree in the intermediate state II.

This transformation may also be obtained by a horizontal strain along the –x/-y direction with smaller lattice constants. Upon a strain or an electric field, the rotated and equivalent state III could be more favorable in energy. As a result, the transformation from I to III can be viewed as both ferroelastic switching and 90 degree ferroelectric switching, which may take place at ambient conditions if the switching barrier is within the desirable range (< 0.15 eV). The symmetrical RS state could still be an intermediate state in a possible switching pathway, and the switching barrier could be moderate. According to our SSNEB calculation, a switching barrier of 0.13 eV/f.u. is obtained for the pathway of 90 degree switching for ZB AgI, which is slightly lower than the 180 degree switching barrier. The identified intermediate state II, however, is a titled metastable structure that is 0.11 eV higher in energy than that of the ground state. As the polarization is switched to –x/-y direction, the lattice constant in -z direction is also swapped with the lattice constant in –x/-y direction, so their ferroelectricity and ferroelasticity are coupled.

For the ferroelastic (90 degree ferroelectric) switching pathway of CuCl displayed in Fig. 3(b), the switching barrier of ~0.15 eV/f.u. is much lowered compared with 180 degree ferroelectric switching. This is plausible as the intermediate state II maintains tetra-coordinate for all atoms, which is energetically more favorable compared with RS state for CuCl. Here, the 180 degree polarization switching can be achieved by repeating 90 degree switching twice, and this pathway is more favorable with a much lower barrier compared with the pathway in Fig. 2(b). As a result, CuCl can be also multiferroic at ambient conditions with coupled ferroelectricity and ferroelasticity.

The epitaxial growth of traditional ferroelectrics like perovskites on silicon is still challenging due to issues such as lattice mismatch, which hinders substitution of silicon-based RAMs by ferroelectric RAMs to be integrated with silicon wafer utilizing a mature silicon process. Having demonstrated a series of ZB multiferroic semiconductors in this study, we illustrate that CuCl and silicon entail similar lattice constants (within 1% lattice mismatch) and structures, as in the case for CuBr and germanium, as well as in the case for ZB AgI and InSb. The lattice similarities have important implication for the epitaxial growth and combination with prevailing semiconductors like silicon, as shown in Fig. S2(a). According to previous experimental reports, both CuBr and AgI are wide-direct-gap semiconductors (band gap in the range of 2.9 - 3.4 eV) with high mobility, while Ge and InSb are both narrow-gap semiconductors (0.2 - 0.7 eV).[6, 33] If the mixing alloy $Cu_{1-x}Br_{1-x}Ge_x$ or $Ag_{1-x}I_{1-x}In_xSb_x$ can be achieved, in view of their similar lattice constants and structures, a chemical multi-junction, as shown in Fig. S2(b), with a broad range of bandgaps might be built, which in principle can overpass the Shockley-Queisser detailed-balance limit of photovoltaic conversion efficiency for mono-bandgap semiconductors.[34] The ferroelectric polarization may enhance the open-circuit voltage and the lifetime of excitons by hindering the recombination of electrons and holes. [35, 36]

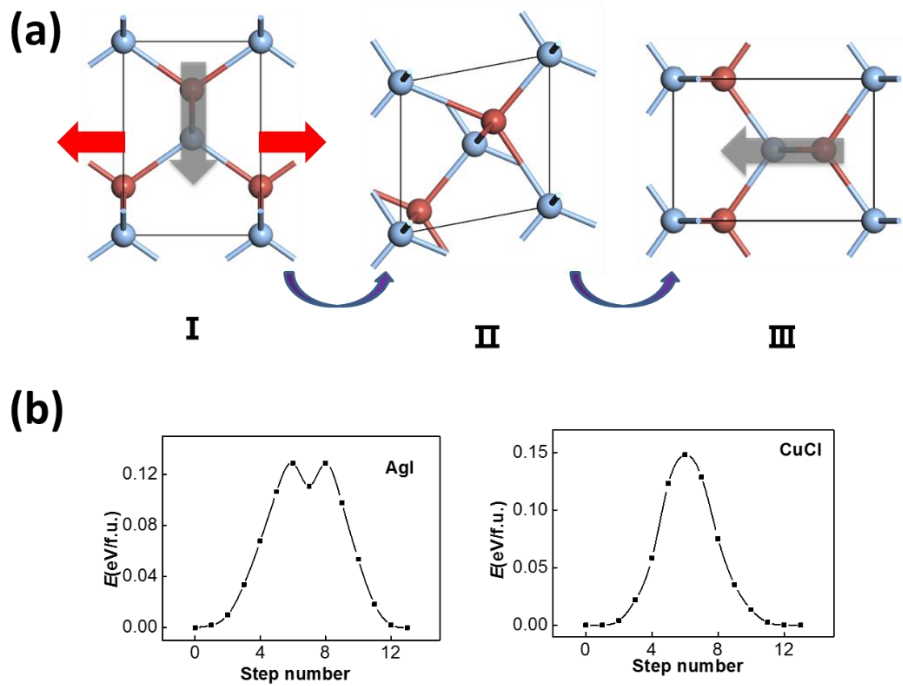

Figure 3. (a) The pathway of ferroelastic switching (90 degree ferroelectric switching) and (b) calculated energy profile based on SSNEB method, for AgI and CuCl. Here the zigzag chains at the initial state (I) are switched by 45 degree at the intermediate state (II), and then by 90 degree in state III.

Table 2. Energy difference between bulk phase and 2D polymorph I/II (ΔI/ΔII), = $E$(2D)-$E$(bulk).

|  |  | AgCl | AgBr | AgI | CuCl | CuBr |
|---|---|---|---|---|---|---|
| **ΔI** | LDA | 0.17 | 0.12 | 0.05 | 0.11 | 0.04 |
| **(eV/f.u.)** | GGA | 0.07 | 0.05 | 0.035 | 0.02 | -0.01 |
| **ΔII** | LDA | 0.24 | 0.15 | 0.08 | 0.13 | 0.07 |
| **(eV/f.u.)** | GGA | 0.13 | 0.08 | 0.05 | -0.01 | 0.00 |

III.     Two-dimensional monolayer polymorphs

We also explore the low-energy monolayer structure of silver/copper monohalides. After an extensive structure search based on CALYPSO program, we obtained two stable structures, I and II, as displayed in Fig. 4(a) for each of CuCl, AgCl and AgBr. The structure I is composed of two buckled layer of honeycomb MX lattice, slightly lower in energy compared with the structure II which is composed of a buckled Cu/Ag monolayer covered by halides at two surfaces. Similar but still different polymorph structures I and II for AgI/CuBr are also obtained, where the structure I is composed of a buckled honeycomb Ag/Cu layer covered by I/Br atoms, with slightly lower energy than the more symmetrical structure II which is composed of a planar square Ag/Cu monolayer covered by I/Br atoms. As listed in Table 2, the energy difference between bulk phase and the 2D polymorphs is around a tenth of eV/f.u., much smaller than that (0.43/0.26 eV/f.u. based on LDA/GGA computation) of the highly ionic crystal of NaCl, or that (0.56/0.44 eV/f.u. based on LDA/GGA computation) of the covalent crystal of ZnS. Fig. 4(c) displays a possible pathway of phase transition from bulk ZB phase to multilayered polymorph II phase, which can be viewed as the rotation of zigzag chains in the red circles. A barrier lower than 0.10 eV/f.u. for such a phase transition of CuCl is predicted based on the SSNEB calculations and LDA method. The negative values for CuCl and CuBr based on GGA computation imply that some 2D polymorphs may be even more favorable in energy compared with ZB phase of bulk structure. Even for LDA results, the 2D polymorph II will become 0.016 eV/f.u. lower in energy compared with ZB phase when half of the Cu atoms in CuCl is substituted by Ni atoms, as shown in Fig. 4(d).

Synthesis of 2D monolayers is likely to be feasible, in view of the small energy difference shown in Table 2. As a comparison, the energy difference between bulk silicon and 2D silicene is notably much larger (0.76/0.65 eV/f.u. by LDA/GGA), and yet 2D silicene has been synthesized in the laboratory. Furthermore, the surfaces covered by halide anions also give rise to oxidation resistance and low cleave energy from layered structure: e.g., the cleavage energy from multilayer to monolayer structure for CuCl I and II are respectively 0.27 and 0.20 J/m$^2$, even lower than that of graphite to graphene (0.3581 J/m$^2$). We have further verified dynamic stability of the 2D monolayers by computing their phonon dispersions, as shown in Figures S3. All the vibration spectra are free of soft modes associated with structural instabilities.

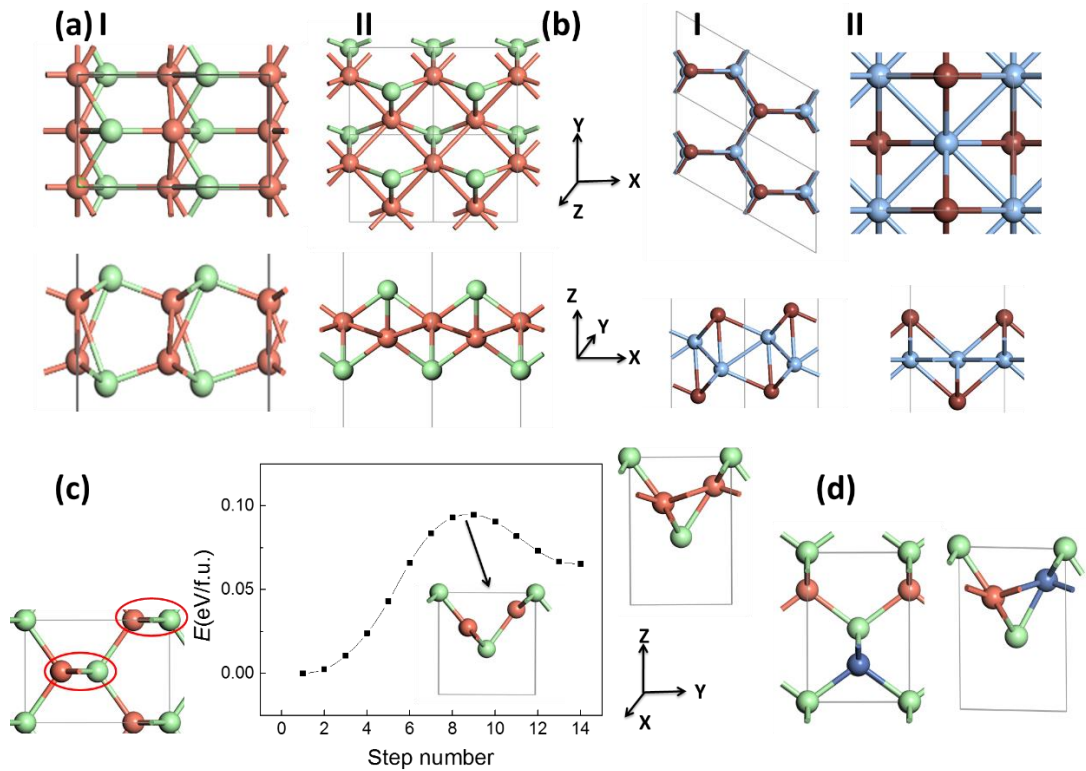

Figure 4. (a,b) Top view (upper panels) and side view (lower panels) of 2D polymorphs, I and II, for (a) CuCl/AgCl/AgBr, and (b) AgI/CuBr. (c) Computed pathway (middle panel) of phase transition for CuCl from bulk ZB (left panel) to multilayered polymorph II (right panel) structure by using LDA method. (d) A comparison of bulk ZB (left panel) and multilayered polymorph II (right panel) structure of $Cu_{0.5}Ni_{0.5}Cl$.

As shown in Fig. 4(a), both 2D structure I and II for CuCl/AgCl/AgBr possess spontaneous in-plane polarization and strain, which may be also multiferroic if they are switchable. Similarly, their polar direction initially along –x axis may be switched to -y axis upon an electric-field, as displayed in Fig. 5(a) and (b), leading to a 90 degree polarization and lattice rotation. This transformation may also be obtained by an in-plane strain along the -y direction, so ferroelastic switching and 90 degree ferroelectric switching are equivalent. According to our SSNEB calculation, a switching barrier of 0.08 eV/f.u. is predicted for the pathway of ferroelastic switching for 2D CuCl polymorph I (Fig. 5(c)), about a half of the barrier height for bulk CuCl, as shown in Fig. 3(b). The 180 degree ferroelectric switching can be obtained by repeating 90 degree switching twice with the same switching barrier, and the switchable polarization of $2.8 \times 10^{-10}$ C/m is even higher than 2D SnS and SnSe. As a result, 2D CuCl I can be also multiferroic at ambient conditions with coupled ferroelectricity and ferroelasticity. The switching barrier for polymorph II is even lower, as displayed in Fig. S4(a), and the switchable polarization is $1.5 \times 10^{-10}$ C/m. The band structures of both 2D polymorphs (Fig. 5(d) and (e)) are highly anisotropic, with distinct effective mass along Γ–X and Γ–Y, so their ferroelastic switching (90 degree ferroelectric switching) can be electrically detected. One may even obtain a 2D triferroics[37] by doping 3d magnetic element. For example, the polymorph II of CuCl becomes ferromagnetic when some of Cu atoms are substituted by Ni atom, as shown in Fig. S4(b). Each Ni atom possesses 1 $\mu_B$ magnetic moment, while the easy axis of magnetism and electrical polarization are both aligned in the same in-plane direction, which can be also switched by 90 degree upon ferroelastic switching. As a result, their ferromagnetism, ferroelectricity and ferroelasticity are all coupled.

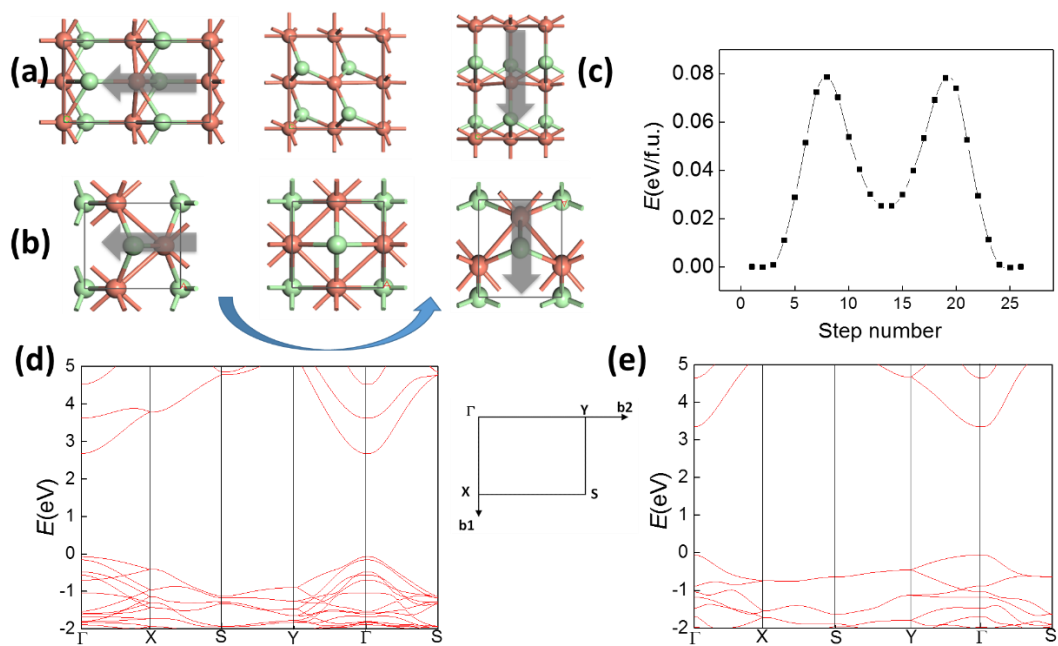

Figure 5. Ferroelastic switching (90 degree ferroelectric switching) of 2D polymorph (a) I and (b) II of CuCl. (c) Calculated pathway of ferroelastic switching for polymorph I. Band structure of polymorph (d) I and (e) II of CuCl are calculated based on the hybrid functional HSE06[38].

In summary, through density-functional theory calculations we demonstrate that the modest barrier for the phase transition between bulk RS and ZB phase may give rise to ferroelectricity and ferroelasticity with relatively low switching barriers at ambient conditions. Some bulk halides even entail similar lattice constants and structures as prevailing semiconductors like silicon, thereby rendering epitaxial growth a high possibility to potentially resolving a major issue of integrating traditional ferroelectrics into silicon-based circuits. Moreover, their stable 2D polymorphs are close or even lower in energy compared with bulk phases, suggesting high likelihood for successful synthesis of these 2D monolayers in future. Some 2D monolayers are also predicted to be 2D multiferroics. As such, their coupled ferroelasticity and ferroelectricity together with anisotropic band structures are highly desired for efficient data reading and writing.

Electronic Supplementary Information is available

**Notes**

The authors declare no competing financial interest.

Acknowledgement

This work is supported by National Natural Science Foundation of China (Nos. 21573084). XCZ is supported by UNL Holland Computing Center.